\documentstyle[epsf]{mn2e}

\DeclareMathAlphabet{\mathbi}{\encodingdefault}{\rmdefault}{\bfdefault}{\itdefault}
\DeclareRobustCommand{\bit}[1]{\ifmmode\mathbi{#1}\else\textbf{\textit{#1}}\fi}
\let\bolditalic=\bit

\newcommand{\be}{\begin{equation}}
\newcommand{\ee}{\end{equation}}

\def\vect#1{{\bolditalic #1}}
\def\thetavec{\vect{\theta}}
\title[Lensing of the OV effect.]
{Observing high redshift galaxy clusters through lensing of the Ostriker-Vishniac effect} 

\author[Diego et al.]  
  {J.M. Diego$^{1}$, D. Herranz$^1$. \\  
   $^1$ IFCA, Instituto de F\'\i sica de Cantabria. Avda. Los Castros s/n. 39005 Santander, Spain. }
\date{Draft version \today}  
  
\pagerange{\pageref{firstpage}--\pageref{lastpage}}  
  
\begin{document}  
\maketitle  
  
\label{firstpage}  
%%%%%%%%%%%%%%%%%%%%%%%%%%%%%%%%%%%%%%%%%%%%%%%%%%%%%%%%%%%%%%%%%%%%%%%%%%%%%%%  
\begin{abstract}  
In this paper we study the possibility of detecting lensing signals in high-resolution  
and high-sensitivity CMB experiments. At scales below 1 arcmin, the CMB background 
is dominated by the Sunyaev-Zel'dovich effect in clusters and by Ostriker-Vishniac 
effect distortions elsewhere. 
Assuming the Sunyaev-Zel'dovich component in clusters can be removed, 
we focus on the Ostriker-Vishniac effect and study the possibility of its detection 
while paying special attention to contaminants, such as instrumental noise and point sources. 
After designing an optimal filter for this particular lensing signal we explore the signal-to-noise 
ratio for different scenarios varying the resolution of the experiment, 
its sensitivity, and the level of contamination due to point sources. 
Our results show that the next generation 
of experiments should be able to do new and exciting science through the lensing effect of the 
Ostriker-Vishniac background. 
\end{abstract}  
%%%%%%%%%%%%%%%%%%%%%%%%%%%%%%%%%%%%%%%%%%%%%%%%%%%%%%%%%%%%%%%%%%%%%%%%%%%%%%%  
%%%%%%%%%%%%%%%%%%%%%%%%%%%%%%%%%%%%%%%%%%%%%%%%%%%%%%%%%%%%%%%%%%%%%%%%%%%%%%%  
%%%%%%%%%%%%%%%%%%%%%%%%%%%%%%%%%%%%%%%%%%%%%%%%%%%  
\section{Introduction}\label{section_introduction}  
%%%%%%%%%%%%%%%%%%%%%%%%%%%%%%%%%%%%%%%%%%%%%%%%%%%  
Future high-resolution CMB or microwave experiments will focus on 
secondary anisotropies such as the Sunyaev-Zel'dovich effect (SZ hereafter, Sunyaev \& Zel'dovich 1972), the 
Ostriker-Vishniac effect (OV hereafter, Ostriker \& Vishniac 1987) effect or the lensing effect. 
Both the SZ effect and the OV effect occur when free electrons interact with CMB photons. During this 
interaction, photons gain energy from the free electrons (this mechanism is known as 
inverse Compton scattering). This gain of energy creates a distortion in the energy 
spectrum of the CMB which is proportional to the temperature of the ionized plasma and its density 
(thermal SZ effect). 
Also, if the plasma cloud is moving with a bulk velocity $v$, this velocity creates an additional 
Doppler shift in the energy of the CMB photons (kinetic SZ). 
Historically the thermal and kinetic SZ effects have been studied in galaxy clusters. 
The OV effect is similar to the kinetic SZ effect but 
focusing on the first free electrons after re-ionization at high redshift. 
The Universe is believed to be fully ionized at redshifts $z < 10$ (e.g Loeb \& Barkana 2001, Kohler, 
Gnedin \& Hamilton 2007). 
At this high redshifts the 
temperature of the ionized medium is too low so the equivalent of the thermal OV can be neglected.
High-resolution, high-sensitivity experiments should be able to see all three effects. The thermal 
SZ effect is the dominant one followed by the kinetic SZ. The thermal SZ effect can be easily identified 
due to its frequency dependence. It should be, in in principle, relatively easy to subtract this signal from 
the maps (see for instance Herranz et al. 2002). The kinetic SZ can be also removed (at least a big part of it) using the spatial distribution of the thermal SZ and the fact that both effects are proportional to the optical depth of the cluster. 
On the other hand, the OV effect spreads over the entire field of view while the SZ effect is concentrated only 
in galaxy clusters. If one observes a small portion of the sky (few arcmin$^2$) it is very possible that 
none or very few clusters will be present in the image and the OV effect may be the dominant anisotropy 
even before subtraction of the SZ effect. Having a high redshift uniform background like the OV effect 
is particularly interesting for lensing studies. If a massive object (for instance a cluster) is in 
the field of view, it will distort the OV effect creating an elongation on the anisotropies in a direction 
perpendicular to the line connecting the cluster and the anisotropy (in the strong lensing regime this elongation 
can be parallel to this direction, i.e. radial arcs). Lensing effect of the OV background opens a new door to study the distribution of high redshift large scale structures. In particular, it may be a powerful tool to study the 
evolution of galaxy clusters detecting proto-clusters at high redshift through their lensing effects. 
Future experiments should look for this signal as it offers a unique way to study the high redshift universe.
In the mm band, there are several experiments currently under development, ALMA  (Wootten 2002), 
ACT (Kosowsky 2003), SPT (Ruhl et al. 2004). Although is very possible that these experiments do not reach the required sensitivity 
they will be important to address some of the issues discussed below, such as contamination by point sources.
Also, high sensitivity, high resolution radio experiments will complement microwave observations 
by looking at the "holes" in the 21-cm line maps (Wang et al. 2006), LOFAR (Rottgering 2003), 
SKA (Carilli 2004). A technique similar to the one presented in this paper could be applied but using the 
21-cm line as a background at even higher redshifts (Koopmans et al. 2004, Zahn \& Zaldarriaga 2006).
While effects like the Sunyaev-Zel'dovich effect have attracted much of the attention, little has 
been done regarding the OV effect 
(Jaffe \& Kamionkowski 1998, Scannapieco 2000, McQuin et al. 2005, Iliev et al. 2006) 
in part due to the difficulties found when simulating the OV effect. One of the main sources of uncertainty 
comes from the poor constraints on the re-ionization history (usually parametrized by the fraction of 
ionized material as a function of redshift $X_e(z)$). Two kinds of constraints have been imposed so far. 
On one hand, studies of the Ly-$\alpha$ forest have found that the Universe is completely reionized from redshift 
$z=0$ to $z \approx 6$. Observations of the Gunn-Peterson trough show that the end of 
the reionization era is at $z \approx 6$ (Fan et al. 2002, Kashikawa et al. 2006). 
On the other hand, WMAP polarization measurements of the cosmic microwave background 
imply that reionization started at $z = 10.9^{+2.7}_{-2.3}$ (Page et al. 2006). This constrain on the beginning 
of reionization was derived assuming that reionization was instantaneous. 
If reionization was a gradual process, then the starting 
point could be as high as $z \approx 20$. The higher the starting point of reionization the denser the Universe was at that point 
and the strongest the OV distortions. Moreover, it is still uncertain how the different scenarios of patchy reionization 
affect the final distribution of OV anisotropies. The resulting uncertainty in the level of OV anisotropies (angular power spectrum) is about one order of magnitude (e.g Santos et al. 2003).

Regarding lensing observations, most of them have focused on optical observations. Particularly interesting are 
strong lensing studies of galaxy clusters since in this case it is possible to have multiple arcs systems coming from 
sources at different redshifts (Broadhurst et al. 1995, 2005a, 2005b, Shapiro et al. 2000, Gladdeers et al. 2002, 
Kneib et al. 2004). Through strong lensing it is possible to estimate the projected mass of a cluster from ten or 
few tens of kpc to about one hundred kpc. Weak lensing studies of galaxy clusters on the other hand allow to extend this 
range to the Mpc scale. Some work has been done to study the lensing distortions on CMB data by galaxy clusters. 
Most of these works focus on distortions of the primary CMB anisotropies or on angular scales larger than 1 arcminute 
(Hirata \& Seljak 2003, Lewis \& King 2006, Hu, De Deo \& Vale 2007). Previous works have tried to 
estimate the lensing effect on smaller scales over secondary anisotropies such as the OV effect 
(Zaldarriaga \& Seljak 1999). These works propose methods to reconstruct the masses of galaxy clusters from their 
lensing signature. Due to its similarity, the reader may find interesting the works by Hu (2001) and  
Maturi et al. (2005a, 2005b). In those papers the authors make use of filters for highlighting the 
lensing signals. In particular, Maturi et al. (2005a, 2005b) defines a filter which is formaly similar 
to the one used in this paper although they apply it to lensed galaxies.

This paper will focus on the possibility of measuring the lensing effect due to high redshift 
clusters through observations of the OV effect and its lensing distortions. We propose a new filter 
in sections 2 and 3, test it with simulated data in section 4 and show the significance of the results for 
various sensitivities, resolutions and point source confusion levels in section 5. 
The same results could be also applied to potential 
observations of the reionization epoch by making high resolution maps of the 21-cm line at those high 
redshifts. We will present a fast and efficient algorithm for highlighting lensing signals in noisy 
data which could be used for estimating the mass of high redshift clusters. 

%%%%%%%%%%%%%%%%%%%%%%%%%%%%%%%%%%%%
\section{Lensing effect}
%%%%%%%%%%%%%%%%%%%%%%%%%%%%%%%%%%%%
The gravitational lensing effect can be quantified by the deflection angle, $\vect{\alpha}$,  
created by the distorting mass (or lens) over the photons. A source which originally 
was located on a position $\vect{\beta}$ in the sky will appear on a different position 
$\vect{\theta}$ at a distance $\vect{\alpha}$ from the original position. 
\begin{equation}  
\vect{\beta} = \thetavec - \vect{\alpha}(\thetavec,m(\thetavec))  
\label{eq_lens}  
\end{equation}  
where $\vect{\alpha}(\thetavec)$ is the deflection angle created by the lens   
which depends on the observed positions, $\thetavec$. 

The deflection angle $\alpha$ at a given position is found by  
integrating the (perpendicular) change in the 3D gravitational potential, $\Phi$, along 
the line of sight;
\begin{equation}  
\alpha = \frac{2}{c^2}\frac{D_{ls}}{D_s D_l} \int \nabla_{\bot} \Phi  dz 
\label{eq_alpha0}  
\end{equation}  
Usually, the projected potential, $\psi$, is used instead of $\Phi$
\begin{equation}
\psi = \frac{2}{c^2}\frac{D_{ls}}{D_s D_l} \int \Phi  dz
\label{eq_alpha1}
\end{equation}
As a function of $\psi$, the deflection angle is just,
\begin{equation}
\alpha = D_l \nabla \psi
\label{eq_alpha2}
\end{equation}
In the above equations $D_{ls}$, $D_l$, and $D_s$ are the angular diameter distances  
from the lens to the source galaxy, the distance from the observer to  
the lens and the distance from the observer to the source galaxy  
respectively. 
From the deflection angle one can easily derive the magnification produced 
by the lens at a given position:
\begin{equation}
\mu^{-1}(\thetavec) = 1 - \frac{\partial\alpha_x}{\partial x} - \frac{\partial\alpha_y}{\partial y} 
                          + \frac{\partial\alpha_x}{\partial x}\frac{\partial\alpha_y}{\partial y} 
                         - \frac{\partial\alpha_x}{\partial y}\frac{\partial\alpha_y}{\partial y} 
\end{equation}
Given the gravitational potential $\psi$,  
the shear is defined in terms of the second partial derivatives of the potential 
$\psi$ (the Hessian of $\psi$):
\begin{eqnarray}
\psi_{ij}		&=&\frac{\partial^2\psi}{\partial\theta_i\partial\theta_j},\\
\gamma_1(\thetavec) 	&=& \frac{1}{2}(\psi_{11} - \psi_{22})
                        = \gamma(\thetavec)\cos[2\varphi],\\
\gamma_2(\thetavec) 	&=& \psi_{12} = \psi_{21} =\gamma(\thetavec)\sin[2\varphi],
\end{eqnarray}
where $\gamma(\thetavec)$ is the amplitude of the shear and $\varphi$ its 
orientation.
The shear can be also expressed in terms of the deflection angle, $\alpha$, yielding
\begin{equation}
\gamma_1 = \frac{1}{2} \left ( \frac{\partial\alpha_x}{\partial x} -
\frac{\partial\alpha_y}{\partial y} \right ),
\label{eq_gamma1}
\end{equation}
\begin{equation}
\gamma_2 = \frac{\partial\alpha_x}{\partial y} = \frac{\partial\alpha_y}{\partial x}.
\label{eq_gamma2}
\end{equation}
The amplitude and orientation of the shear are given by 
\begin{equation}
\gamma = \sqrt{\gamma_1^2 + \gamma_2^2},
\end{equation}
\begin{equation}
\varphi = \frac{1}{2} \mathrm{atan}(\frac{\gamma_2}{\gamma_1}).
\end{equation}
\begin{figure}  
   \epsfysize=4.cm   
   \begin{minipage}{\epsfysize}\epsffile{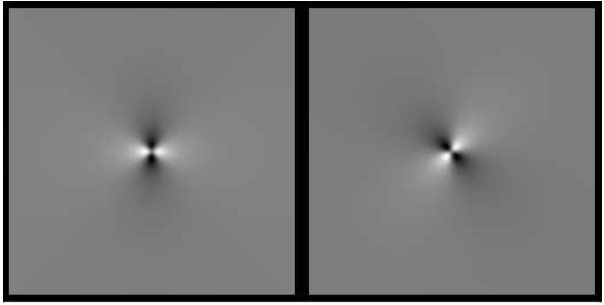}\end{minipage}  
   \caption{Shear distortion maps $\gamma_1$ (left) and $\gamma_2$ (right) for a cluster 
            at the center (the cluster has arbitrary mass and redshift).  
            The images have been smoothed for representation purposes. 
           }  
   \label{fig_stokes1}  
\end{figure}  
In figure \ref{fig_stokes1} we show the distribution of $\gamma_1$ and $\gamma_2$ for a cluster with 
a Navarro, Frenk and White (1997) profile (NFW profile hereafter).  
When measuring shear distortions, the reduced shear $g$ is measured instead;
\begin{equation}
g = \frac{\gamma}{1-\kappa}
\label{eq_g}
\end{equation}
with the convergence $\kappa$ defined by;
\begin{equation}
\kappa = \frac{1}{2} \left ( \frac{\partial\alpha_x}{\partial x} +
\frac{\partial\alpha_y}{\partial y} \right ),
\label{eq_kappa}
\end{equation}
Since $\gamma$ has two components, the reduced shear will have two components as well, 
$g_1$ and $g_2$. 
In figure \ref{fig_stokes2} we show the expected reduced shear signal ($g_1$ and $g_2$) signal for 
a cluster (of arbitrary mass and redshift). The $g_1$ and $g_2$ maps show a distinctive 
pattern which divides the space in two regions. 
The division between the central region and the rest occurs when the convergence $\kappa=1$. 
This coincides with the tangential critical curve (Einstein radius) or region where 
the magnification is maximum. 
When crossing this line in a radial direction the nature of the lensing arcs change from tangential to radial 
(or viceversa). This can be appreciated by a change in the sign of both $g_1$ and $g_2$. 
The ring shaped distribution in $g_1$ and $g_2$ will lose its symmetry if the lens is not symmetric. 
One expects the division between the two zones to follow the critical curve, defined by the region where 
the projected surface mass density equals the critical density. 

\subsection{Quantifying shear distortions in CMB maps}.
%%%%%%%%%%%%%%%%%%%%%%%%%%%%%%%%%%%%%%%%%%%%%%%%%%%%%%%
Weak lensing analysis historically has been applied to optical images where a large number of 
background sources (galaxies) is observed and their ellipticities measured. By averaging the 
ellipticities of many galaxies ($\approx 100$) over a given area ($\approx 1$ arcminute$^2$), 
an estimate of the reduced shear, $g$, is obtained in that area. Repeating the same process over 
different contiguous areas it is possible to give an estimate of the gravitational potential responsible for 
the measured average ellipticities. CMB maps are different because one lacks individual images (galaxies) 
and therefore is not possible to define constrained images with well defined borders for which an 
ellipticity (or orientation) can be derived. Instead, one deals with continuous fields (the OV effect 
or field $f$) having a typical anisotropy scale of 10-60 arcseconds. For this part of the discussion we 
will assume the field $F$ consists only of pure OV effect signal (with lensing distortions). In a real 
experiment one will have also systematic noise and small scale contaminants such as point sources. 
We will discuss these later. The anisotropies of the OV effect have intrinsic ellipticities which 
introduce random correlated shears on these scales of 10-60 arcseconds. 
One expects the ellipticities of contiguous anisotropies (by contiguous we mean separated more 
$\sim$ 1 arcmin) not to be correlated so the mean orientation of a sufficiently large area should 
average to 0 (as it happens with the standard weak lensing of background galaxies). Given the scale of 
the OV effect this means we should average areas of many arcminutes in order to 
reduce the intrinsic shear scatter of the OV effect. This poses a huge problem for lensing 
analysis since the typical scale of lensing signals is a few arcmin for distant moderate mass 
clusters. Averaging the shear over many arcminutes would wash out any lensing signal from these 
clusters. We have to find a way to solve this problem. One option is to reduce the natural s
cale of the OV effect (or field $f$). 
This can be done by working with the curvature map, $C_f$, of the field $f$ instead. 
\begin{figure}  
   \epsfysize=4.cm   
   \begin{minipage}{\epsfysize}\epsffile{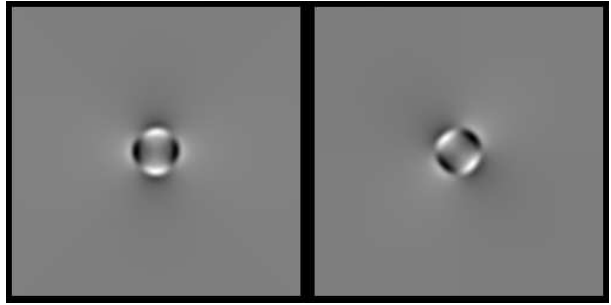}\end{minipage}  
   \caption{Reduced shear maps $g_1$ (left) and $g_2$ (right) for the same cluster 
            of figure \ref{fig_stokes1}. The images have been smoothed for representation purposes. 
           }  
   \label{fig_stokes2}  
\end{figure}  
The curvature will reduce the large scale component of the OV effect and will enhance small scale 
fluctuations. Lensing signals will be minimally affected by this since the lensing stretches 
the anisotropies, therefore adding small scale features which will be enhanced when we compute 
the curvature of the map. The curvature of $f$ can be defined as;
\begin{equation}
C_f = \frac{\partial ^2f}{\partial x^2} + \frac{\partial ^2f}{\partial y^2}  
\label{eq_Cf}
\end{equation}
We can anticipate here that when dealing with noise data, the curvature would be dominated 
by the pixel noise. This problem can be significantly reduced by filtering out the noise 
(for instance with a Gaussian filter) before calculating the curvature. We will discuss this point 
in more detail later.

The orientations (or shear) of the curvature field, $C_f$, can be constructed from the first 
derivatives of the curvature field. 
Based on these derivatives it is possible to construct quantities similar to the Stokes parameters $Q$ and $U$.
\begin{equation}
Q_c = \frac{1}{2}((\frac{\partial C_f}{\partial x})^2 - (\frac{\partial C_f}{\partial y})^2) = \frac{1}{2}(C_x'^2 - C_y'^2)
\label{eq_Qc}
\end{equation}
\begin{equation}
U_c = \frac{\partial C_f}{\partial x}\frac{\partial C_f}{\partial y} = C_x'C_y'
\label{eq_Uc}
\end{equation}
An associated intensity $I_c$ can be defined as;
\begin{equation}
I_c = \sqrt{Q_c^2 + U_c^2}
\end{equation} 
For the shake of simplicity, from now on we will drop the suffix $c$ from $Q$, $U$ and $I$.
We should notice the correspondence between the definition of the quantities 
($Q$, $U$) and ($\gamma_1$, $\gamma_2$). Like in the case of $\gamma_1$, $Q$ is positive
when the shear is aligned vertically while $Q$ is negative when the shear is aligned horizontally. 
In the case of $U$ the situation is similar but with a 45 degrees rotation 
(like in the case of $\gamma_2$). The $U$ parameter is maximum when the features are aligned 
over the 45 degrees direction and minimum when they are aligned over the -45 degree direction. \\ 

For speed purposes it is convenient to compute the curvature map $C_f$ 
in Fourier space.
\begin{equation}
\tilde{C}_f = -k^2\tilde{f}
\end{equation}
where tilde denotes Fourier transforms, $k=(k_x,k_y)$ is the Fourier k-mode 
(or wave vector) and $f$ is the original data.

Also, the first derivatives of the curvature map can be computed in Fourier space.
\begin{equation}
\tilde{C_x'} = ik_x\tilde{C_f}
\end{equation}
\begin{equation}
\tilde{C_y'} = ik_y\tilde{C_f}
\end{equation}
The Stokes parameter fields $Q$ and $U$ are obtained after Fourier transforming back 
to real space the complex maps $\tilde{C_x'}$ and $\tilde{C_y'}$ and substituting them 
in equations (\ref{eq_Qc}) and (\ref{eq_Uc}). 
Computing the derivatives this way has the advantage of being faster but it has the inconvenience of 
being affected by border effects (due to periodical boundary conditions in the Fourier transform). 
We can remove border effects by neglecting the areas near the image borders. 
 
\section{Optimal filters for lensing in CMB data}
%%%%%%%%%%%%%%%%%%%%%%%%%%%%%%%%%%%%%%%%%%%%%%%%%%
The chances of detecting a weak signal embedded in a noisy background
(in this case, the shear distortion created by a cluster in the CMB)
can be greatly enhanced by using specifically tailored linear
filters. The linear filters that maximize the signal to noise ratio of
a given signal are known as \emph{linear matched filters} (often
referred to as \emph{optimal filters}).  A matched filter is obtained
by correlating a known signal, or template, with the given data in
order to detect the presence of the template in the data. In presence
of pure white noise, the shape of the filter is the same as the shape
of the signal to be detected. However, when the noise has spatial
correlations the noise covariance matrix (or, equivalently, its power
spectrum) must be taken into account. The general form of the matched
filter is proportional in Fourier space to the template of the signal
over the power spectrum of the background

Matched filters have been widely used in engineering and data analysis
for decades (Turin et al. 1960) and in many areas of
Astronomy, including detection of galaxy clusters in X-ray surveys
(Vikhlinin et al. 1995, Stewart et al. 2006), in three-dimensional
redshift surveys (Kepner et al. 1999), detection of
extragalactic point sources in CMB maps (L\'opez-Caniego et al. 2006) 
and detecion of themal Sunyaev-Zel'dovich
effect (Herranz et al. 2005, Melin et al. 2006). Besides, matched filters have been used
for the detection of galaxy clusters through weak lensing in
(Maturi et al. 2005a,2005b). Bernstein \& Jarvis (2002) presented optimal techniques 
to measure the weak gravitational shear from images of galaxies.
In these works a matched or optimal filters for the shear were constructed. 
In this work we will focus instead in designing
matched filters for the Stokes parameters $Q$ and $U$ of the curvature field.

After decomposing the curvature map into its $Q$ and $U$ 
components (using the Fourier technique) we can apply an appropriate filter to highlight 
the expected lensing signal from a galaxy cluster. 
Optimal filters are generally defined (in Fourier space) as the signal (theoretical) we want to detect 
divided by the power spectrum of the background.  
The theoretical lensing signal can be computed numerically for a given profile calculating the 
deflection angles $\alpha_x$ and $\alpha_y$ and then the reduced shear $g = (g_1,g_2)$. The power spectrum 
of the background can be obtained directly from the data (in our case the power spectrum of the $Q$ and 
$U$ maps). The negative is that we need to assume a cluster profile in order to compute $g$. This problem 
can be partially solved if we adopt a modified version of the definition of $g$. We will assume the reduced 
shear can be well described by;
\begin{equation}
\hat{g}_1 = \gamma_1\times(1 + \kappa)
\label{eq_g1}
\end{equation}
\begin{equation}
\hat{g}_2 = \gamma_2\times(1 + \kappa)
\label{eq_g2}
\end{equation}
We use the hat notation to distinguish between the formal definition of $g_1$ and $g_2$ (see equation \ref{eq_g}). 
In the weak lensing limit where $\kappa << 1$ equations (\ref{eq_g1}) and (\ref{eq_g2}) are similar to equation 
(\ref{eq_g}) (to first order in $\kappa$). 
The above equations will fail in describing the correct shear where $\kappa \approx 1$ or $\kappa > 1$. 
This range corresponds to the strong lensing case. Our new definition for $g_1$ and $g_2$ will have no negative 
consequences if the strong lensing part of the data is omitted. This is exactly what we are going to do. 
By omitting the strong lensing part of the data we avoid two important problems. On one hand, the Einstein ring 
shown in figure \ref{fig_stokes2} disappears since with the new definition there is no change in sign when 
$\kappa \approx 1$. This will make our results less insensitive (and therefore more robust) to any assumption 
about the specific profile of the cluster. 
On the other hand, by excluding the central region of the cluster we avoid some of the most 
serious contaminants, in particular the kinetic Sunyaev-Zel'dovich effect due to internal bulk motions of the 
gas in the cluster and also mm or sub-mm extragalactic sources. The kinetic SZ can not be removed with multifrequency 
observations (like the thermal SZ) since it does not depend on the frequency. Point sources within the galaxy 
cluster will be unresolved at the resolutions we are considering here (fwhm $> 5$ arcsec) and they can be very  
close to each other making the use of matched filters even harder due to their blending.
One way to avoid the central region of the cluster is by means of a damping function. 
\begin{figure}  
   \epsfysize=4.cm   
   \begin{minipage}{\epsfysize}\epsffile{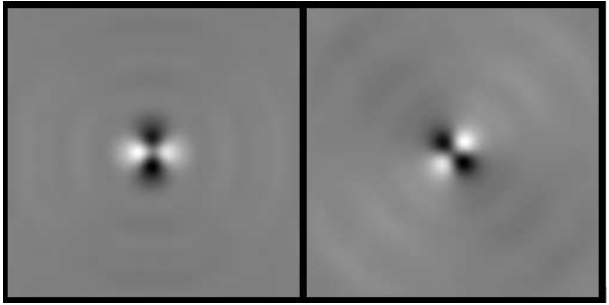}\end{minipage}  
   \caption{Optimal filters $F_Q$ (left) and $F_U$ (right) obtained for one of our simulations.
            Note how the intensity goes to 0 at the center as a consequence of the damping function.
            The oscillations are typical of matched filters.
           }  
   \label{fig_Filter}  
\end{figure}  
This leads us to the final form of the theoretical signal that we will assume 
for constructing the optimal filter.
\begin{equation}
\hat{g}_1 = (1 - G(\sigma))^2\times \gamma_1 \times (1 + \kappa)   
\label{eq_h1}
\end{equation}
\begin{equation}
\hat{g}_2 = (1 - G(\sigma))^2\times \gamma_2 \times (1 + \kappa)   
\label{eq_h2}
\end{equation}
where $G(\sigma)$ is a Gaussian function with dispersion $\sigma$.
The above definition introduces a scale 
$\sigma$. The role of this scale is somehow similar to the Einstein radius but with the difference that we only 
require that $\sigma$ is of the order of the Einstein radius (a factor 2 difference does not make a big difference 
as we will see below). 
Although we are not going to discuss about alternative choices for $\hat{g}_1$ and $\hat{g}_2$ we want to note 
that simplified versions of the reduced shear gave also good results when applied to our our simulated data. 
In particular, due to its simplicity we can emphasize the following one;
\begin{equation}
\hat{g}_1 = (1 - G(\sigma))^2\times \frac{x^2-y^2}{x^2 + y^2}
\label{eq_h1}
\end{equation}
\begin{equation}
\hat{g}_2 = (1 - G(\sigma))^2\times \frac{2xy}{x^2 + y^2}
\label{eq_h2}
\end{equation}
where $x$ and $y$ are the distance to the center of the cluster. 
This form reminds one of the kernel proposed by Squires \& Kaiser 1996) 
although the use of that kernel and our filter is different in each case.

In figure \ref{fig_Filter} we show the filter for $\sigma=0.3$ arcmin (field of view of 12 arcmin).
If the contaminants can be accounted for and removed, then 
the strong lensing signal could be measured as well. In this case, filters like $g_1$ and $g_2$ 
(see figure \ref{fig_stokes2}), may be more appropriate. 

Once the expected signal has been defined, the optimal filters (one for $Q$ and one for $U$) are 
constructed in a standard way.
\begin{equation}
\tilde{F}_Q = \frac{\tilde{\hat{g}}_1(k)\times \tilde{b}(k)}{P_Q(k)}
\label{eq_Fq}
\end{equation}
\begin{equation}
\tilde{F}_U = \frac{\tilde{\hat{g}}_2(k)\times \tilde{b}(k)}{P_U(k)}
\label{eq_Fq}
\end{equation}
Where as above tilde denotes Fourier transform. The term $\tilde{b}(k)$ is the Fourier transform of the 
kernel used to filter out the noise. We will discuss this factor in more detail later but we prefer to include it 
here for completion. We use a Gaussian kernel for $b$ with a scale larger than the resolution of the experiment. 
$P_Q(k)$ and $P_U(k)$ are the power spectrum of the $Q$ and $U$ maps respectively (original data $->$ filter noise 
$->$ curvature $->$  $1{st}$ derivatives $->$ $Q$ and $U$). 
A very similar philosophy has been followed in the past to define optimal filters for weak lensing 
optical surveys (Maturi et al. 2005b). 
Although there are some differences, our filter and theirs are closely related. 
Some differences are the use of the damping function in our filter to reduce cluster contamination 
or the use of the total (background) power instead of the noise power spectrum in their definition 
of the filter. Alos, in their work the authors aopply their filter to simulated optical data (with 
background galaxies acting as the source plane for lensing).

%%%%%%%%%%%%%%%%%%%%%%%
\section{Simulations}
\begin{figure}  
   \epsfysize=6.cm   
   \begin{minipage}{\epsfysize}\epsffile{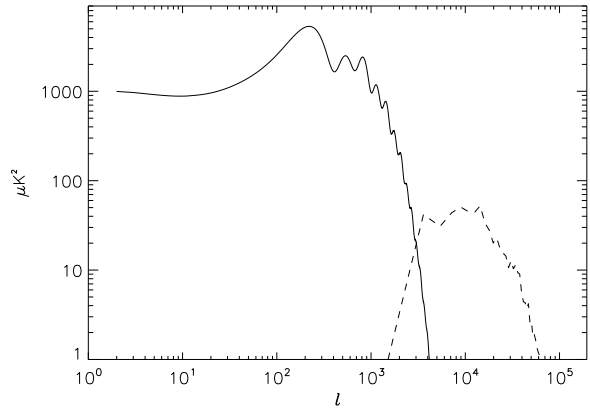}\end{minipage}  
   \caption{Power spectrum of the simulated OV effect (dashed line) compared with 
            the power spectrum of the CMB (solid line).
           }  
   \label{fig_Cl}  
\end{figure}

\begin{figure*}  
   \begin{flushleft}
   \epsfysize=5.8cm   
%   \epsfysize=12.5cm   
%   \begin{minipage}{\epsfysize}\epsffile{M4D14_Sigma4Dm6_256pix_6panels.ps}\end{minipage}  
   \begin{minipage}{\epsfysize}\epsffile{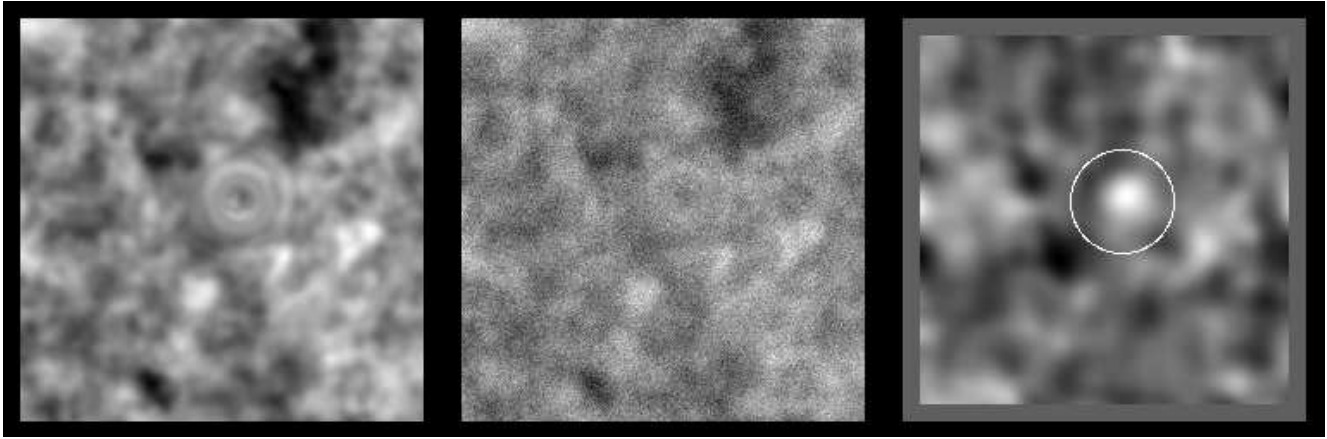}\end{minipage}  
   \caption{The panels show the original simulated OV effect in a $12\times12$ arcmin$^2$ field 
            (top-left). There is a cluster at redshift $z=1$ and with mass $M = 4\times 10^{14} h^{-1} M_{\odot}$ near 
            the center. The cluster potential distorts the space around it creating distinctive lensing patterns. 
            The middle panel shows the same map after being observed with an experiment having a  
            resolution of 8.43 arcsec fwhm and a noise level of $3 \mu$K per pixel (or equivalently $0.15 \mu$K 
            per arcmin$^2$.) The right panel shows the final result obtained after combining the optimally 
            filtered Q and U maps. This cluster is at the limit of detectability (SNR $\approx 5$)
           }  
   \label{fig_data_result}  
   \end{flushleft}
\end{figure*}

%%%%%%%%%%%%%%%%%%%%%%%
Testing the algorithm with simulations is essential, not only to prove 
its feasibility but also to identify its failures and weaknesses. 
We first simulate the OV effect when CMB photons cross 
the first newly re-ionized regions at high redshift. 
After the universe recombined at redshift $z \approx 1100$,  
it underwent a period know as the {\it dark ages} where the small density perturbations (of the order 
of $10{-5}$ at redshift z=1000) started to grow and form larger structures. During this period the Universe was 
neutral and transparent to CMB photons. At some point around redshift $z \approx 20$ 
it is believed that the first stars and/or quasars re-ionized the medium around them (UV radiation). 
As the times passed this newly re-ionized zones started to grow and merge forming larger and larger 
ionized areas. Free electrons in ionized regions would re-scatter a fraction ($\approx 10\%$) of the 
CMB photons (inverse Compton scattering). 
Due to the low energy of CMB photons the scattering is elastic 
and isotropic (the scattered photon can leave the interaction in any direction). However, the CMB 
photon suffers a Doppler shift in its spectrum due to the electrons motion. 
The distortion in the spectrum of the CMB photons is then proportional to the bulk velocity of the 
free electrons and their optical depth, $\tau$.
\begin{equation}
\frac{\Delta T}{T} = -\frac{v_r}{c} \tau
\end{equation}
with $v_r$ the radial component of the electron bulk velocity (or projection along the line of sight).
\begin{figure}  
   \epsfysize=6.cm   
   \begin{minipage}{\epsfysize}\epsffile{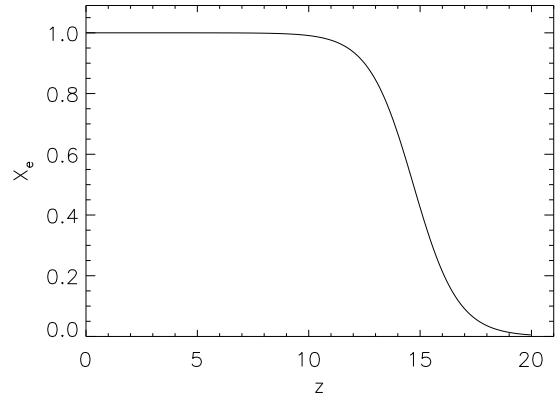}\end{minipage}  
   \caption{Fraction of ionized gas as a function of redshift.
           }  
   \label{fig_reion}  
\end{figure}  
As mentioned in the introduction, the OV effect is similar to the kinetic SZ. However, the later 
is commonly used when we refer to hot-ionized plasma in galaxy clusters while the OV effect focuses 
on ionized regions in much larger scales. In the introduction we mentioned also that the equivalent 
of the thermal OV effect can be considered negligible since the temperature of the intergalactic medium 
(IGM) is much smaller than 
in galaxy clusters ($T_{IGM} \approx 10^4$ K, Haehnelt \& Steinmetz 1998, Hui \& Haiman 2003).
Although the Universe is also ionized at low redshift, this distortion is expected to be dominated by the 
high redshift Universe because it is denser by a factor $(1+z)^3$. 
The interaction between the free electrons and the CMB photons will be then more 
likely at high redshifts. To simulate the OV effect we use the public code GADGET-2 
(Springer et al. 2005). 
We make 6 different simulations of 50 Mpc (comovil) cubes each and with more than 2 million 
particles per cube. The simulation is started at redshift $z \approx 40$ until $z \approx 3$. 
When projected, the cubes always have a scale larger than 20 arcmin for any redshift.
We adopted a stacking procedure to project the simulations along the line of sight. 
We stack 45 projections of $12\times12$ arcmin each (50 Mpc deep). 
This give us a range in redshift between $z_{min} = 4.6$ and $z_{max} = 20.7$ (assuming $\Omega_m = 0.27$ and 
$\Lambda = 0.3$). This range of redshifts guarantees that we do not miss the beginning or end of 
the reionization epoch ($5 < z < 20$). 
Since we have only 6 independent simulations and in order to avoid correlated
repetitions of the same projected zone, we need to rotate and/or take different patches or 
quadrants (of 12 arcmin each). We do not impose continuity conditions in the overlapping 
zone of the boxes but we expect this to be a minor systematic effect given the dimensions of the cubes. 
We assumed a patchy reionization and with a reionization history starting at redshift $z \approx 20$ and leading 
to a fully ionized Universe at redshift $z \approx 10$ (see figure \ref{fig_reion}). This ionization history is 
compatible with the upper limit of the latest WMAP constraints 
(Page et al. 2006). 
We assumed that high density regions were ionized first (Sokasian et al. 2003).  
Taking a homogeneous reionization instead reduces the OV power at small scales by a 
factor 2 or 3.
The OV effect is computed by projecting along the line of sight the radial velocity times the 
optical depth (which depends on the reionization history). The final map shows features at small 
and large angular scales. The RMS peaks at scales between 20 arcseconds and 1 arcmin 
(see figure \ref{fig_Cl}) with values of RMS $\approx 5 \mu$K. 
When compared with the CMB power spectrum, the OV spectrum dominates at scales smaller than 2 arcmin (assuming the constribution from instrumental noise, point sources and galaxy clusters can be removed).
This results lay on the high end but are still comparable to previous estimations of this effect (Jaffe \& Kamionkowski 1998).
Our simulations are simplistic in the sense that we do not take into consideration all the physical processes 
involved in the ionization of the IGM. Instead we reduce all the physics to the ionized fraction. However, 
these simulations will suffice for the purpose of this paper.
The final map of the OV effect is used as a background template for computing the lensing 
effect of clusters at redshift $z = 1$.
We use the public package WSLAP for the computation of the lensing effect (Diego et al. 2004 etc)
The resulting map is convolved with a Gaussian kernel to simulate the experimental beam and then is pixelized. 
We take 3 pixels per resolution element (fwhm). 
White Gaussian noise is added afterwards to account for instrumental noise. 
We will consider different noise levels and resolutions (4.2, 8.5 and 17 arcsec fwhm). More details on the 
resolution and noise levels will be given later. As discussed above, in order to reduce the intrinsic 
shear of the OV effect we work with the curvature map instead. However, before computing the curvature the noise 
has to be filtered out since otherwise pixel to pixel variations due to the instrumental noise would 
dominate the curvature. The curvature map is then used to derive the $Q$ and $U$ maps of the data. These two maps 
are then optimally filtered and added to render the final result. In figure \ref{fig_data_result} we show an example 
for a cluster at redshift $z=1$ and mass $M = 4\times 10^{14} h^{-1} M_{\odot}$. The noise level was $3 \mu$K per pixel 
and the beam of the experiment had a $fwhm = 8.43$ arcsec. The combination of Q and U filters produce a clear 
detection with a significance of $5\sigma$. 
The RMS $\sigma$ is computed in the area outside the white circle in the 
right panel. The borders have been ignored because of systematic effects due to computations in Fourier space.
It is interesting to note how the filter does not highlight features that at first can look like lensing effects. 
For instance, there is a round structure, similar to the lensing signal, on the top left side of the image. 
This structure is not enhanced at the same level as the real lensing signal.
%%%%%%%%%%%%%%%%%%%%
\section{Results}  
%%%%%%%%%%%%%%%%%%%%  
The simulated data has to be pre-processed as we mention earlier. The $Q$ and $U$ maps are computed over the 
curvature map of the data. However, the local curvature would be completely dominated by the instrumental 
noise if we do not attempt to filter it out first. For this purpose we apply a Gaussian filter with a 
dispersion $\sigma_g$ which depends on the level of noise. For white Gaussian noise the RMS  
of the filtered noise decreases as $1/\sigma_g$. Depending on the noise level we can choose the appropriate 
$\sigma_g$ so the noise level of the filtered map does not dominate the curvature of OV effect.
%Filtered noise levels below $\approx$ 30 microK per arcmin$^2$ are required to be able to detect lensing 
%signals of moderate mass distant clusters. For instance, an experiment with a resolution of 8.5 arcseconds 
%and noise level of $0.2 \mu$K per arcmin$^2$ (or similarly $4 \mu$K per pixel) will decrease the noise level 
%down to $\approx 20$ microK per arcmin$^2$ if the noise is filtered with a Gaussian having  
%$\sigma_g \approx 30$ arcseconds. 

The results will depend on the characteristics of the experiment. 
In particular its sensitivity and resolution but it will also depend crucially on other contaminants 
like the Sunyaev-Zel'dovich effect in galaxy clusters (thermal and kinetic) and more importantly mm and sub-mm 
sources. The thermal Sunyaev-Zel'dovich effect can be removed if several frequencies are available since 
this effect follows a well established frequency dependence. The kinetic SZ on the other hand is frequency 
independent but it is proportional to the thermal SZ (at least the component due to cluster bulk motion) 
so it can be also partially removed. However, the removal might not be perfect and some residuals may be left 
specially in the central part of the cluster where energetic phenomena can create wave fronts or fast internal 
motions of the gas difficult to extract (Mathis et al. 2005). However, as mentioned earlier, our strategy 
to deal with these contaminants is neglecting the central regions of galaxy clusters by using a damping function 
in the center of the filter. Our most worrying source of contamination will be point sources but before studying 
their effect we will focus on the other aspects affecting the results.

\subsection{Effects of the scale $\sigma$}
%-----------------------------------------
First we study the effect of the damping function discussed above. The role of this scale can be identified as 
the mass of the cluster. Increasing the scale $\sigma$ changes the optimal filters in a way that makes bigger 
scales more relevant. It also has an effect on the {\it peaks} of the signal. 
A large $\sigma$ removes more contributions coming from the central part of a cluster but it also removes 
contributions coming from small scale fluctuations in the curvature map. Hence, if there are still significant 
small scale contributions to the curvature coming from instance from instrumental noise or point sources, the damping 
function helps to remove these signals. It is important to note that the effect of the smoothing function (used to 
filter out the noise before computing the curvature) is similar to the damping function. When a smoothing kernel 
is applied over the theoretical shear, $\hat{g}_1$ and $\hat{g}_2$, it softens the peak compensating the positive and 
negative contributions at the center. One could then be tempted to remove the damping function from the definition of 
the optimal filter but this would have a negative effect on high signal-to-noise experiments 
where the noise does not suppose a serious threat. In this case, filtering out of the noise is not necessary but 
still a damping of the filter in its center is recommended to remove contaminants and reduce contributions from 
isolated peaks in the curvature map. In order to evaluate the effect of $\sigma$ we will consider 
experiments with low noise (otherwise the effects of noise filtering can dominate the effect of the damping function). 
We find that a change in $\sigma$ of almost an order of magnitude renders basically the same number of detections 
(SNR $>$ 5). Values of $\sigma$ much smaller than the expected Einstein radius of a cluster produce results which 
are very sensitive to isolated peaks in the curvature map. On the other hand, values of $\sigma$ much bigger than 
a typical Einstein radius produce maps which are too diluted. Values for $\sigma$ in the range $20-60$ arcseconds 
produce in general satisfactory results. The small end will be more suitable for distant lower mass clusters while the 
higher end is more appropriate for detecting nearer massive clusters.

\subsection{Effects of instrumental noise}
\begin{figure}  
   \epsfysize=6.cm   
   \begin{minipage}{\epsfysize}\epsffile{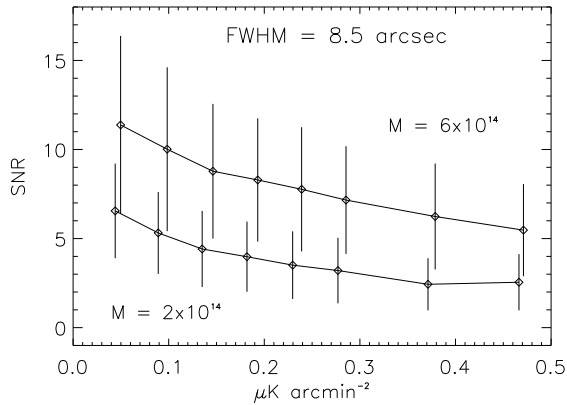}\end{minipage}  
   \caption{SNR as a function of noise level for two cluster mass. The clusters 
            are at redshift $z=1$ and masses are given in units of $h^{-1} M_{\odot}$. Error bars correspond 
            to $1 \sigma$.
           }  
   \label{fig_SNR_noise}  
\end{figure}

\begin{figure}  
   \epsfysize=6.cm   
   \begin{minipage}{\epsfysize}\epsffile{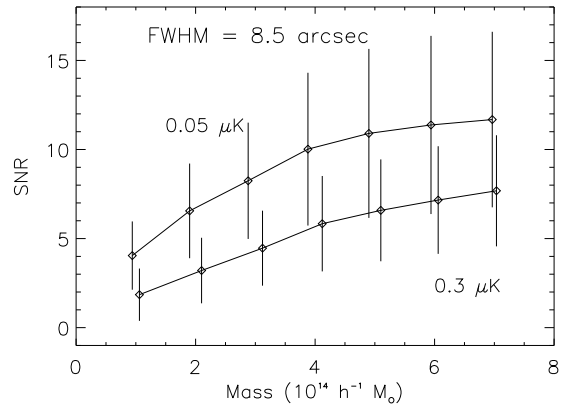}\end{minipage}  
   \caption{SNR as a function of cluster mass for two levels of instrumental noise. 
            The upper curve is for a noise level of $0.05 \mu$K per arcminute$^2$ and 
            the lower curve is for a noise level of $0.3 \mu$K per arcminute$^2$. 
            The cluster is assumed to be at redshift $z=1$ and the experiment has a 
            resolution of 8.5 arcsec per beam. 
           }  
   \label{fig_SNR_mass}  
\end{figure}

%-----------------------------------------
We assume that instrumental noise can be well described by a Gaussian distribution with dispersion 
$\sigma_N$. We also assume that the noise is uncorrelated (white noise). In order to compare sensitivities 
the noise is normally expressed as $\mu$K per arcmin$^2$. We will follow the same notation and give our noise 
levels in the same units although in some cases it will be useful to express the noise in $\mu$K per pixel.
At the scales of the resolution elements of our simulated data (4.2, 8.5 and 17 arcsec) 
the signal we are simulating (the OV effect) has an RMS of a few $\mu$K.  The noise 
level should be of the same order or below the signal level itself (otherwise the experiment would be 
noise dominated and we would not see any signal at all). This means that the noise per pixel should not be 
more than a few $\mu$K. This noise level per pixel translated into $\mu$K per arcmin$^2$ gives 
$\approx 0.1 \mu$K per arcmin$^2$. This sensitivity is, for today's standards, about only one order of magnitude 
better than current experiments. In order to make a statistical analysis we make 100 simulations for 
each case varying the noise (but keeping $\sigma_N$) and position of the cluster in each one. After taking different noise levels and applying the formalism discussed above 
we compute the SNR (for the 100 cases) as the ratio between the filtered signal at the central position of the cluster and the RMS computed 
in the area excluding the cluster. The result as a function of noise level can be seen in figure \ref{fig_SNR_noise}.
We plot SNR and its dispersion (from the 100 simulations) for two cluster masses at redshift $z=1$. In this result, the simulated experiment 
had a resolution of $fwhm = 8.5$ arcsec. 
From the plot we see that for noise levels under $0.1 \mu$K per arcmin$^2$ it is possible to 
detect a significant fraction of clusters down to masses of $M = 2 \times 10^{14} h^{-1} M_{\odot}$. At noise levels 
of $0.5 \mu$K per arcmin$^2$ approximately half the clusters with masses  $M = 6 \times 10^{14} h^{-1} M_{\odot}$ 
can be detected at these resolutions (in the absence of other contaminants). This mass is expected to be 
the largest cluster mass at $z=1$. On the other hand, most of the low mass clusters ($M < 2 \times 10^{14} 
h^{-1} M_{\odot}$) will be undetected with this noise level and resolution. 
A similar result is shown in figure \ref{fig_SNR_mass}  but this time fixing the noise level and varying the mass. 
Two noise levels are shown in the plot. The clusters are again assumed to be at $z=1$. 
Both plots show larger variations in the SNR with increasing mass (even for smaller noise levels). The error bars 
depend on three factors. First the noise level, second the background OV and third the fact that we use an optimal 
filter which depends on the background. The smaller contributions to the variation in the SNR are the noise 
level and the change in background which enters the definition of the optimal filter. Most of the variation 
in the SNR comes from the OV background behind the cluster. A background rich in features and big 
changes in intensity will produce sharper arcs which will be more easily identified in the curvature map. On the 
other hand, if the OV background behind the cluster is soft and poor in features, the lensing effect will show 
wide gradients in temperature rather than well defined arcs. This second case will be much more difficult to 
detect through the shear distortions. Bigger clusters have a larger lensing cross section. Hence they are 
sensitive to a wider range of scales and consequently their SNR can change more.

\subsection{Effects of spatial resolution}
\begin{figure}  
   \epsfysize=6.cm   
   \begin{minipage}{\epsfysize}\epsffile{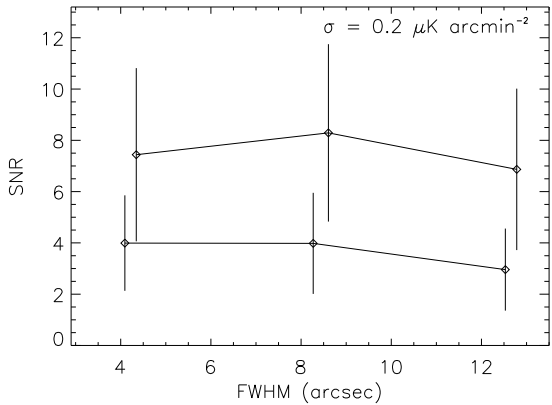}\end{minipage}  
   \caption{SNR as a function of the experiment resolution. The noise level per arcminute is 
            the same for the three simulated data sets. The upper curve shows the SNR for 
            a clusters with $M = 5 10^{14} h^{-1} M_{\odot}$ while the bottom curve is 
            for clusters with $M = 2 10^{14} h^{-1} M_{\odot}$. Also shown are the 
            $1-\sigma$ error bars.
           }  
   \label{fig_SNR_fwhm}  
\end{figure}

%-----------------------------------------
Spatial resolution plays an important role on the detectability of this effect. 
%First, a experiment 
%with poor resolution (for example $fwhm > 30$ arcsec) will dilute the small scale anisotropies of the OV 
%effect. On the other hand, high resolution experiments (for example $fwhm < 5$ arcsec) ...
As discussed in the last paragraph in the last subsection the detectability of a cluster depends very much on 
the OV background behind the cluster. If the background is rich in structure the cluster will be detected 
more easily while if the background has small variations in intensity it will be more difficult to measure 
shear distortions. The beam of an experiment has two negative effects in lensing. On one hand it rounds the 
anisotropies (if the beam is symmetric) reducing the shear of the anisotropy. On the other hand, it softens the 
anisotropies, removing variations in intensity smaller than the beam resolution. The ideal experiment will 
have a beam size smaller than the range where the RMS of the OV anisotropies is maximum ($20-60$ arcsec).
In this case we will consider three beam sizes; 4.2, 8.5 and 17 arcsec. Larger beams will dilute 
the OV anisotropies hence making harder to detection of lensing signals. Also, larger beams makes the issue of  
point source confusion a more serious one. On the other hand a small beam 
is technologically more challenging since a larger collecting area is required and the sensitivity has to be 
increased accordingly with the resolution.  
In figure \ref{fig_SNR_fwhm} we show the effect of changing the spatial resolution of the experiment while keeping 
its sensitivity (note that the pixel per noise must be different in each case to maintain the sensitivity). Again we 
show the results for two different masses. Our results show that, in the range of resolutions relevant for the OV 
effect, the beam resolution of the experiment does not seem to have a large impact on the final results (although 
larger beams would start to reduce the SNR drastically).   
The explanation for why the resolution does not seem to affect the result is because we are 
setting the noise level per arcmin$^2$ the same for the three cases (resolutions). The same noise level per 
arcmin$^2$ implies however a higher noise level per pixel in the higher resolution experiment. This means that  
filtering out the noise has to be done with a Gaussian which in absolute terms is equivalent in the three cases. 
We could say that at the end the effective resolution is determined more by the noise level than by the 
experimental beam. Since in figure \ref{fig_SNR_fwhm} the three experiments have the same noise level, the result 
is basically the same in all three cases.

\subsection{Effects of point sources}
\begin{figure}  
   \epsfysize=6.cm 
   \begin{minipage}{\epsfysize}\epsffile{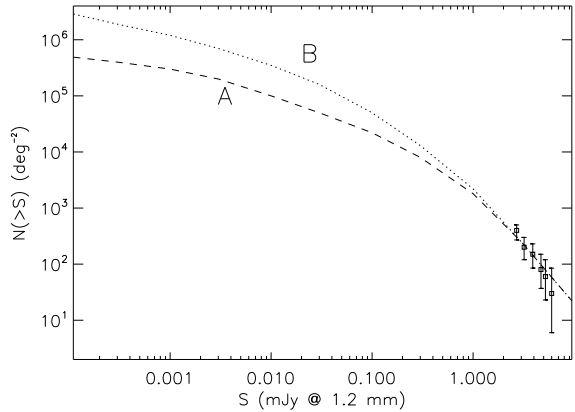}\end{minipage}  
   \caption{Integrated number counts for the two models used to simulate the point sources. 
            The lack of data below $S \sim 1$ mJy at these frequencies makes necessary to 
            consider extrapolations. Data points are observations made with MAMBO 
            at 1.2 mm (Greve et al. 2004). 
            }  
   \label{fig_PS_AB}  
\end{figure}

\begin{figure}  
   \epsfysize=6.cm   
   \begin{minipage}{\epsfysize}\epsffile{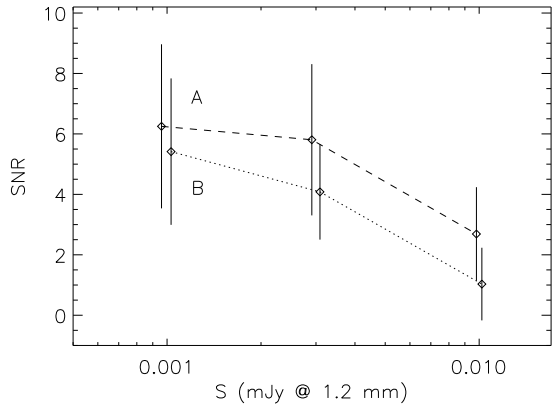}\end{minipage}  
   \caption{Signal-to-noise ratio as a function of the exclusion limit for point sources. 
            The two curves correspond to the two models A and B of figure \ref{fig_PS_AB}
            }  
   \label{fig_SNR_PS_AB}  
\end{figure}

%------------------------------------
Finally we explore the most serious source of systematic error; galactic sources. To asses their importance we 
simulate a population of point sources assuming they are observed at 1.2 mm (250 GHz). 
Due to the lack of observations at these wavelengths, little is known about the number counts in this range. 
Sub-mm and mm observations are still in their earlier stages. 
One of today's most precise measurements of the number counts of galaxies at 1.2 mm wavelengths comes from the 
MAMBO experiment (Smail, Ivison \& Blain 1997, Hugues et al. 1998, Barger, Cowie \& Sanders 1999, 
Bertoldi et al. 2000, Eales et al. 2000). A deep survey with MAMBO (Greve et al. 2004) revealed that number counts 
at 1.2 mm where about one order of magnitude less than at 850 $\mu$m (SCUBA number counts). 
This drop in the number counts is partially due to the fact that mm sources are generally dimmer 
when wavelength increases. In particular, combining MAMBO and SCUBA sources one finds that the ratio of fluxes 
$S_{850\mu m}/S_{1200\mu m}$ is about a factor three (Greve et al. 2004). At even longer wavelengths one expects 
that number counts are even smaller. This suggests that observing at frequencies around 100 GHz might be optimal to 
reduce the negative effects of point sources.  
To simulate the point sources at 1.2 mm we use the MAMBO number counts of 
Greve et al. (2004). In the same paper the authors compare their number counts 
with models having different galaxy evolution. These models are compatible with both 
SCUBA and MAMBO number counts.
Due to lack of information below 1 mJy (at 250 GHz or equivalently 1.2 mm), 
we extrapolate the model to lower fluxes to account for the dimmer sources. We consider two different 
extrapolations that are compatible with the observed number counts. In figure \ref{fig_PS_AB} we show the integrated 
number counts of our two models. 
The flattening of the integrated number counts at low fluxes in figure \ref{fig_PS_AB} is predicted by 
simulations (Fardal et al. 2001).
In our calculations, we assume that we can remove all sources above a given flux. Sources need to be subtracted 
before filtering out the noise and computing the curvature of the map. In this paper we assume the sources can 
be removed perfectly. This is an optimistic assumption as subtraction algorithms always leave a small residual. 
It is beyond the scope of this paper to study the effects of these residuals here and we will focus only 
on the negative effects of the sources which can not be removed. 
For the shake of simplicity, our sources are simulated assuming that they are not affected by lensing 
distortions. This is approximation is valid if the angular resolution of the experiment is substantially 
larger than the typical size of galaxies. 
With the resolutions considered in this work (fwhm $\approx 8$ arcsec) all our sources will be unresolved. 
We should notice that by doing this approximation we are being conservative since sources might actually help 
to highlight lenses since they could introduce shears in the maps. However, we will not consider this case  in 
this work. In order to check the effect of point sources we will add a population of sources up to a given 
flux cut. All sources above this flux cut are assumed to be removed from the map. 
Figure \ref{fig_SNR_PS_AB}, summarizes our findings for three different cuts in flux. Our result 
predicts that sources are a major contaminant and that they need to be removed down to the microJy 
level (at 1.2 mm). This will be a serious issue which needs to be investigated further but not in this paper. 
For instance, multifrequency observations might be required to reduce the level of contamination of mm sources.

%%%%%%%%%%%%%%%%%%%%%%%%%%%%%%%%%%
\section{Discussion of the results}  
%%%%%%%%%%%%%%%%%%%%%%%%%%%%%%%%%%
Our simulations show how an experiment with a resolution of $fwhm \approx 8$ arcsec and sensitivity of 
$\sigma_N \approx 0.2 \mu$K arcmin$^{-2}$ could detect high-redshift galaxy clusters 
through their lensing signature on the OV anisotropies. Experiments like ALMA will be soon near this limit.  
Current experiments like the Atacama 25-m telescope (ACT) can reach sensitivities of 
$\approx 2 \mu$K with beams of 1.7 arcminute or less at mm frequencies (Kosowsky 2003). 
The main limiting factor 
to detect distant clusters through OV-lensing will be our ability to subtract point sources. Our simulations show 
how one needs to subtract sources down to a few microJy (at $\lambda=1.2$ mm) 
in order to observe the OV anisotropies and their 
lensing distortions. In the flux range 1-3 microJy (or 2-6 $\mu$K per beam) one expects nearly one source per beam 
($fwhm \approx 8$ arcsec). Although the microJy level may seem unreachable, 
we have to notice that sources above 3 microJy flux can be seen as 
$\approx 5 \sigma$ fluctuations with the sensitivities considered in this paper ($0.2 \mu$k arcmin$^{-2}$). 
Therefore it should be straightforward to subtract them. Also, even for lower flux sources, the fact that 
their flux vary with frequency (while the OV effect does not) can be used to develop subtraction algorithms 
for multifrequency maps (Herranz et al. 2002). These algorithms can also use information about the shape of 
sources (which at these resolutions will be unresolved) to define a matched filter. 
Another limiting factor will be the sensitivity of the instrument. We estimate that sensitivities below 
$0.2 \mu$k arcmin$^{-2}$ are necessary in order to reliably detect lensing signals. This level of sensitivity is about 
one order of magnitude better than current experiments. ALMA for instance expects sensitivities (in the continuum) 
of about 40 microJy with only 60 second integration in the frequency band 125-163 GHz (Wootten 2002). Longer integration 
times would increase the sensitivity thus allowing to observe the OV anisotropies (6 ks would increase the 
sensitivity one order of magnitude to 4 microJy). 
Regarding the spatial resolution of the experiment we found 
that this is not as important as long as it is better than $\approx 10$ arcsec. Smaller beams though, will allow 
a better subtraction of point sources. These kind of resolutions can be obtained with large single dish telescopes 
(LMT) or interferometers (ALMA).  
Achieving these sensitivities and resolutions will open a new window for 
cluster studies, allowing the detection (and mass estimation) of high redshift clusters even before they can 
be observed by other means. 
This is particularly interesting since at high redshift the geometric factor in the lensing signal decreases 
more slowly than at lower redshifts. For example, increasing the redshift of the cluster from $z=1$ to 
$z=4$ decreases the amplitude of the lensing signal only by a factor 2. 
  
\subsection{Alternative choices for the filter}
%%%%%%%%%%%%%%%%%%%%%%%%%%%%%%%%%%%%%%%%%%%%%
It is interesting to notice that there is still room for improvement. 
For instance, we have presented results for an optimal filter but other choices 
could render similar results. An interesting alternative is the operator defined by 
Kaiser and Squires (1993). In equation (2.1.12) of that paper they derive the Fourier 
transform of the mass surface density as; 
\begin{equation}
\tilde{\Sigma}(\vect{k}) = - \chi_i(\vect{k}) \tilde{e}_i(\vect{k})
\label{eq_Kaiser}
\end{equation}
where $\vect{k}$ is the wavevector in Fourier space,  
$\tilde{e}(\vect{k})$ is the Fourier transform 
of the ellipticity and $\chi(\vect{k})$ is the operator.
\begin{equation}
\chi_1(\vect{k}) =\frac{k_1^2-k_2^2}{k_1^2 + k_2^2}
\end{equation} 
\begin{equation}
\chi_2(\vect{k}) =\frac{2k_1k_2}{k_1^2 + k_2^2}
\end{equation} 
Taking Q and U as our estimates of the ellipticities and applying equation 
\ref{eq_Kaiser} renders results very similar to the ones presented in this work after  
the Q and U maps are averaged over scales of 20-30 arcsec. We should notice that while 
our filter acts as a convolution over the Q and U maps, the Kaiser and Squires operator 
acts as a combination of second derivatives of the Q and U maps. 

%%%%%%%%%%%%%%%%%%%%%
\section{Conclusions}  
%%%%%%%%%%%%%%%%%%%%%
The OV effect can create anisotropies of a few $\mu$K on scales of $10-30$ arcseconds.
A galaxy cluster can distort an OV effect background creating a distinctive lensing signature. This 
distortion produces a shear of the OV anisotropies which can be highlighted with an appropriate filter. 
We have seen how an optimal filter can be built from the Stokes parameters of the curvature map of the image. 
We test the potentiality of the filter to highlight lensing signals with simulated data. The simulated data includes 
different levels of instrumental noise, angular resolution, and point source contamination. We study how the 
instrument characteristics and point source contamination affect the result, finding that point sources might 
be the most serious problem and that it will be necessary to subtract them down to the microJy level 
at mm wavelengths. 
The effort to go down to this level should be rewarding. These type of studies would open a new and 
exciting window to explore cluster formation, since it would allow to detect and 
estimate the mass of high redshift cluster through their shear distortions. 
A more careful analysis of the lensed images could predict the mass of the cluster with better accuracy. 
For instance, the Einstein radius could be easily identified in many cases since a continuous background 
guarantees that there will always be photons originating at the caustics.

%%%%%%%%%%%%%%%%%%%%%%%%%%%  
\section{Acknowledgments}  
%%%%%%%%%%%%%%%%%%%%%%%%%%%  
The authors would like to thank E. Mart\'inez-Gonz\'alez, P. Vielva, G. Bernstein, B. Jain, Elizabeth E. Brait for useful comments. 
%Finally, we would like to thank 
%the referee for his/her dedication and very careful review of our paper.

%%\newpage  
  
%%%%%%%%%%%%%%%%%%%%%%%%%%%%%%%%%%%%%%%%%%%%%%%%%%%%%%%%%%%%%%%%%%%%%%  

%%%%%%%%%%%%%%%%%%%%%%%%%%%%%%%%%%%%%%%%%%%%%%%%%%%%%%%%%%%%%%%%%%%%%%%  

\bsp  
\label{lastpage}  
\end{document}